\documentstyle[12pt]{article}

\newcommand{\bb}{\begin{equation}}
\newcommand{\ee}{\end{equation}}
\newcommand{\ba}{\begin{array}}
\newcommand{\ea}{\end{array}}
\newcommand{\beqa}{\begin{eqnarray}}
\newcommand{\eeqa}{\end{eqnarray}}

\newcommand{\MS}{$\bar{MS}$ }
\newcommand{\lnM}{\ln{\frac{M^2}{\mu^2}}}
\newcommand{\lnm}{\ln{\frac{m^2}{\mu^2}}}

\newcommand{\ep}{effective potential }
\newcommand{\cc}{coupling constant }
\newcommand{\lamb}{\overline{\lambda}}
\newcommand{\lamB}{$\bar{\lambda}$ }
\newcommand{\yyb}{\overline{y}^2}
\newcommand{\yyB}{$\bar{y}^2$ }
\newcommand{\ggsb}{\overline{g}_s^2}
\newcommand{\ggsB}{$\bar{g}_s^2$ }
\newcommand{\Mb}{\overline{M}}
\newcommand{\MMb}{\overline{M}^2}
\newcommand{\mb}{\overline{m}}
\newcommand{\mmb}{\overline{m}^2}
\newcommand{\Ib}{\overline{I}}
\newcommand{\zv}{\overline{\zeta}_v}
\newcommand{\rh}{\frac{\overline{\lambda}}{\overline{y}^2}}
\newcommand{\br}{\beta_{\rho} }
\newcommand{\gwwb}{\overline{g}_2^2}
\newcommand{\gwb}{\overline{g}_1^2}

\begin{document}

\title{On the Lower Bound for the Higgs Boson Mass}
\author{R.S.Willey\thanks{\it Physics Dept, University of Pittsburgh,
Pittsburgh PA 15260;    E-Mail:willey@vms.cis.pitt.edu}}

\date{December 3,1995}

\maketitle

\begin{center}
Abstract\\
\end{center}

We provide an alternative derivation of a lower bound on the mass of the
Higgs boson which is somewhat simpler and more direct than the derivation
based on the effective potential. For one TeV cutoff, the result is the
same. For high scale cutoff, the lower bound is  increased by slightly
more than the expected uncertainty in the calculation.
\vspace{.3in}

\vspace{.3in}

\clearpage

    For the history of the subject we refer to reviews \cite{Sher} and quote
only two recent papers containing the latest refinements of the conventional
approach \cite{AI}. Once the experimental lower bound on the top-quark mass
exceeded $80\,  GeV$, attention shifted from the original Linde-Weinberg bound,
based on the properties of the one-loop effective potential for small $\phi$,
close to the minimum, to the large $\phi$ behavior of the \ep , as determined
by renormalization group (RG) considerations.
\[
V_{eff}(\phi)=\frac{1}{4}\lamb (t)(\xi (t)\phi )^4
\]
Here, $\xi (t)$ is the anomalous dimension factor, and $\lambda (t)$ is the
\MS running \cc .
\[
t=\ln{\frac{\phi}{M_0}}, \hspace{.5in} \frac{d \lamb (t)}{dt}=
\beta_{\lambda} (\lamb (t),\overline{g}(t))
\]
It is then argued that vacuum stability requires \lamB $(t) > 0$ up to
some high scale, $\phi \sim M_{GUT}$,\,or$M_{Pl}$,\,($M_0 \sim M_Z$,\,or $m_t$,
or
$246 GeV$).To implement this condition, one has to know (or approximate) the
$\beta -$ function, integrate the RG differential equations starting from some
initial values,\lamB $(0)$,\,$\bar{g}^2(0)$, and relate the smallest acceptable
\lamB $(0)$ to a physical Higgs mass.

     We do appreciate this calculation, but questions may be raised about
the perturbative nature, the scale ambiguity, and the conceptual basis.The
exact $\beta-$ function is not known, so one integrates the one- loop  $\beta-$
functions
to get the ''RG improved one- loop \ep''. Since the coupling is not
asymptotically free, the large t behavior of \lamB $(t)$ is not known. The best
that can be
(and is) achieved is perturbative self-consistency. For the indicated
\lamB $(0)$, integrating the one- loop $\beta-$ function gives \lamB
 \nolinebreak{(t)}
which remains perturbative up to the high scales considered. The minimum
\lamB $(0)$ depends on the minimum scale $M_0$ above. The physical Higgs
mass does not. So there is some scale ambiguity. On the third point, we
largely repeat the remarks of \cite{Lin}. If one thinks of a nonperturbative
formulation of the vacuum stability problem, in particular, a lattice
formulation; the large $\phi$ behavior of the \ep is not the point. On the
lattice, the exact \ep is well defined and convex. The condition for vacuum
stability is simply that the bare quartic coupling constant must be positive
($\lambda_{bare}> 0$). Then the lower bound on the Higgs mass is just the
smallest output Higgs mass from the Monte Carlo simulation as one runs through
the space of bare parameters in the broken symmetry phase.

    We present a new derivation of the Higgs mass lower bound. It is also of
the perturbative RG variety, and so also subject to the first concern above;
but we have organized the calculation in a way which minimizes the scale
ambiguity and makes no explicit reference to the \ep.
 The essential input is that one is perturbing
about the correct vacuum. A neccessary condition for this is that the vev of
the (shifted) field  be zero,
order by order in perturbation theory, and the renormalized mass
squared of the shifted field be positive.

   We start by computing the relation between the perturbative pole mass and
the \MS mass, for both the Higgs boson and the t-quark. The
relation follows from the perurbative definition of the pole mass,
\bb
0 =\overline{D}^{-1}(M^{*^2}) = M^{*^2}-\overline{M}^2 -Re\,\overline{\Sigma}
(M^{*^2})      \label{defpole}
\ee
In this equation, $\bar{D}(q^2)$,\, and $\bar{\Sigma}(q^2)$ are the two-
point Green Function and self-energy function, renormalized according to the
\MS prescription. $M^*$ is the perturbative pole mass.The result is
\bb
M^{*^2}=\MMb\{1+\lamb (3 \Ib _{00}(M^{*^2})+ 9 \Ib _{\Mb \Mb}(M^{*^2})) + N_c
\yyb(\frac{M^{*^2}}{\MMb} -4 \frac{\mmb}{\MMb})\Ib_{\mb\mb}(M^{*^2}) + 2(
\zv -1)\}   \label{MM}
\ee
m is the t-quark mass, and y is the t-quark Yukawa coupling ($m = \frac{yv}{
\sqrt{2}}$).
The contributions from the electroweak gauge sector, proportional to $g_2,g_1$,
have also been calculated, but are not written out here.
they will be included below. The term $\zv -1$ comes from a finite shift
of the vev required in the \MS scheme to enforce $<\hat{H}>=0$ through one-
loop order. It will cancel out of the ratio computed below, so we do not have
to give its value here. \cite{BW}$ \bar{I}_{ab}$ is the dimensionally
regularized \MS scalar one-loop two-point integral.
\bb
 \ba{c}
  \Ib_{ab}(q^2)=[\mu^{4-d} i\int \frac{d^d l}{(2\pi)^d}\frac{1}{(l^2 -a^2)
   ((l-q)^2 -b^2)}]_{_{\overline{MS}}}  \\
  = \frac{1}{16\pi^2}[\ln{\frac{ab}{\mu^2}} + \int_{0}^{1} dx \ln{\frac{a^2 x +
  b^2 (1-x) -q^2 x(1-x)}{ab}}]
 \ea
 \label{Iab}
\ee
Then (\ref{MM}) is
\bb
 M^{*^2}=\MMb\{1 + \frac{\lamb}{16 \pi^2}[12 \lnM -24 + 3\sqrt{3}\pi] + N_c
 \frac{\yyb}{16 \pi^2}(1-\frac{4}{r^2})[\lnm + f(r)] + 2(\zv -1)\}
  \label{MM2}
\ee
where
\[
 r=\frac{M}{m}, \hspace{.5in} f(r)=-2+2\sqrt{\frac{4-r^2}{r^2}}\arctan{
 \sqrt{\frac{r^2}{4-r^2}}}
\]
The corresponding calculation for the t-quark gives  \cite{BW1}
\bb
 m^{*^2}=\mmb \{1+\frac{\yyb}{16 \pi^2}[\frac{3}{2}\lnm +\Delta (r)] +
 \frac{\ggsb}{16 \pi^2}C_F(8-6\lnm) + 2(\zv -1)\}  \label{mm}
\ee
where
\[
 \Delta(r)= -4+\frac{r^2}{2}+(\frac{3}{2}r^2 -\frac{1}{4}r^4)\ln{r^2}+\frac
 {r}{2}(4-r^2)^{\frac{3}{2}}\arctan{\sqrt{\frac{4-r^2}{r^2}}}
\]
We take the ratio of (\ref{MM}) to (\ref{mm}) and expand to one-loop order.
\bb
 \ba{c}
  \frac{M^{*^2}}{m^{*^2}}=\frac{\MMb}{\mmb}\{1+\frac{\lamb}{16 \pi^2}[12\lnM -
  24+3\sqrt{3}\pi] + \frac{\yyb}{16 \pi^2}[N_c(1-\frac{4}{r^2})(\lnm +f(r))-
  \frac{3}{2}\lnm -\Delta (r)] \\
   +\frac{\ggsb}{16 \pi^2}C_F(6\lnm -8) + g_2^2,g_1^2 \, terms + 2-loop\}
 \ea
  \label{rat1}
\ee
The $\zv -1$ terms, which also contain explicit dependence on $\ln{\mu^2}$,
have cancelled out.
A necessary condition for the \MS perturbation calculations to be defined
in the broken symmetry phase is that $\bar{M}^2,\bar{m}^2$ be positive.
Since the ratio of pole masses is positive, (\ref{rat1}) satisfies the
requirement perturbatively, for $\mu$ around the weak scale. For large
$\mu^2$, one has to provide a RG treatment of the large logarithms, just
as in the conventional calculation involving the \ep.
In the broken symmetry phase, one can define the renormalized coupling
constants such that the relation
\bb
 \frac{M^2}{m^2}=4\frac{\lambda}{y^2}  \label{ccrat}
\ee
is exact when all the quantities are either ''star'' (on-shell renormalization
scheme) or ''bar''(\MS renormalization scheme)\cite{BW}. Thus, not all
quantities in (\ref{rat1}) can be varied independently as functions of $\mu$.
We focus particularly on the $\frac{4}{r^2}$ multiplying the $\ln{\frac{
m^2}{\mu^2}}$. Tracing its origin to a ratio of \MS masses in (\ref{MM}),
we use (\ref{ccrat}) to replace the ratio of \MS masses by a ratio  of \MS
coupling constants. To leading (one-loop) order, the scale dependence of
the ratio of \MS masses is determined by the coefficients of the explicit
$\ln{\mu^2}$ terms in (\ref{rat1}). For the other masses in (\ref{rat1}), the
difference between ''star'' and ''bar'' is higher order (combined with
explicitly two-loop effects),as is the implicit $\mu$ dependence of the
''bar'' coupling constants. After these observations, and reinstating the
$g_2^2,g_1^2$ terms, differentiating (\ref{rat1}), we obtain
\bb
 \ba{c}
  \mu \frac{d}{d\mu}(\rh)=\frac{\yyb}{16 \pi^2}[24 (\rh)^2 +(2N_c -3 +12 C_F
  \frac{\ggsb}{\yyb})\rh -2 N_c  \\
  -(\frac{9}{2}\frac{\gwwb}{\yyb}+\frac{1}{6}\frac{\gwb}{\yyb})\rh
  +\frac{3}{4}\frac{\overline{g}_2^4}{\overline{y}^4}
  +\frac{3}{8}(\frac{\gwwb +\gwb}{\yyb})^2] + 2-loop
 \ea
  \label{deriv}
\ee

    We now give a sequence of estimates, of increasing refinement, of the
ratio $\frac{M_h^2}{m_t^2}$. Let $\frac{\lamb}{\yyb}=\rho$. (By (\ref{ccrat}),
$\rho = \frac{r^2}{4}$). Let the right hand side of (\ref{deriv}) be denoted
$\beta_{\rho}$. Because of the $-2 N_c$ term in (\ref{deriv}),there is a
 critical value of $\rho$ below which $\br$ becomes negative. And if the
   starting value
of $\rho$ is below this value, as $\rho$ decreases the derivative becomes
more negative, driving $\rho$ negative, unless some higher order effect
intervenes. We will return to this possibility, but as our zeroth order
estimate we take the critical value of $\rho$ for which one-loop $\br$,
evaluated with weak scale coupling constants, is zero. We take $g_s^2 =
1.366$ ($\alpha_s(174) = .109$) and $y^2=0.9990$ ($m_t=174,v=246.2$) and
neglect the $g_2^2,g_1^2$ contributions. Then
$\rho_c = 0.2019$, which gives $(\frac{\bar{M}^2}{\bar{m}^2})^2=0.8076$,
or $\bar{M}_c = 156$, for $\bar{m}=174$. If we include the contribution from
the electroweak gauge couplings, $g_2^2,g_1^2$, the corresponding results are
$\rho_c = 0.2068$, and $\bar{M}_c = 158$, a one percent shift.

The first refinement is to convert back from $(\frac{\bar{M}^2}{\bar{m}^2})_c$
 to the ratio of squared perturbative pole masses by (\ref{rat1}). Note
that precisely for $\rho = \rho_c$, all of the $\ln{\mu^2}$ terms in
 (\ref{rat1})
 cancel, so there is no explicit dependence on $\mu$ in this
correction. The result is
\bb
 (\frac{M^{*^2}}{m^{*^2}})_c = (\frac{\MMb}{\mmb})_c (1-0.101)  \label{cor1}
\ee
which gives $M_h^*\geq 148$.

   We now turn to the effect of the running of the \MS coupling constants,
which appear as coefficients in (\ref{deriv}),and the dependence of the
lower bound on the cutoff (maximum value of $\frac{\mu}{\mu_0}$). One has
to integrate coupled RG equations for five independent ''coupling
constants'', \ggsB ,$\bar{g}_2^2$,$\bar{g}_1^2$,\yyB,$\rho $.
 Let $t=\ln{\frac{\mu}{\mu_0}}$.

\bb
 \ba{c}
  \frac{d}{dt}\ggsb = -\frac{1}{16 \pi^2}(22-\frac{4}{3}N_g)\overline{g}_s^4
      \\
    \frac{d}{dt}\overline{g}_2^2 = -\frac{1}{16 \pi^2}[\frac{44}{3}-\frac{8}
     {3}N_g -\frac{1}{3}N_d]\overline{g}_2^4  \\
    \frac{d}{dt}\overline{g}_1^2 = \frac{1}{16 \pi^2}[\frac{40}{9}N_g +\frac
     {1}{3}N_d]\overline{g}_1^4  \\
    \frac{d}{dt}\yyb = \frac{1}{16 \pi^2}[(3+2 N_c)\overline{y}^4
     -12 C_F \ggsb \yyb -\frac{9}{2}\overline{g}_2^2\yyb -\frac{17}{6}
    \overline{g}_1^2\yyb]  \\
    \frac{d}{dt}(\rho)=\frac{\yyb}{16 \pi^2}[24 \rho^2 +(2N_c -3 +12 C_F
  \frac{\ggsb}{\yyb})\rho -2 N_c  \\
  -(\frac{9}{2}\frac{\gwwb}{\yyb}+\frac{1}{6}\frac{\gwb}{\yyb})\rho
  +\frac{3}{4}\frac{\overline{g}_2^4}{\overline{y}^4}
  +\frac{3}{8}(\frac{\gwwb +\gwb}{\yyb})^2]
 \ea
  \label{5diff}
\ee
The first three equations are integrated trivially. If we neglect the $\bar
{g}_2,\bar{g}_1$ contributions to the $\bar{y}$ running, that equation can
also be integrated analytically. But if one runs up to high scales, the
electroweak gauge couplings become of same order as the QCD coupling constant;
so we use NDSolve from Mathematica to provide an interpolating function
solution for \yyB which is subtituted into the \ggsB equation, which is again
integrated numerically by NDSolve.

  If we run up to the Planck scale ($m_{pl}\approx 10^{19}$ GeV, $t\approx 41$)
the smallest starting $\rho (0)$ which does not lead to $\rho (t)$ falling
through zero before $t = 41$ is $0.2022$, which is only slightly different
from the critcal value required to make the derivative zero at the weak scale.
At the other extreme, if we only require the equations of the standard model
to be consistent up to order of one TeV, the  value of t which corresponds to
one Tev depends on the choice of $\mu_0$. We choose $\mu_0 = m_t$ as the most
natural choice for relating $\bar{m}$ to $m^*$. For $m=174$, this corresponds
to $t_{max}= 1.75$. Then the smallest starting $\rho(0)$ which does not lead
to $\rho (t)$ falling through zero before $t = 1.75$ is $0.0520$. Making the
connection back to the ratio of pole masses by (\ref{rat1}), we obtain the
final result of this approach

\bb
 \ba{c}
   m_t^* = 174, \hspace{.2in} \overline{\alpha}_s(m_t) = .109  \\
      \mu_{max} \approx m_{Pl} \hspace{.5in} M_h \geq 148 \, (141,135) \\
      \mu_{max} \approx 1 TeV  \hspace{.5in} M_h \geq 72  \, (72)
 \ea
  \label{result}
\ee
The numbers in parentheses are the corresponding results in the conventional
aproach \cite{AI}. The lower bound obtained in the present approach is
slightly higher in the high scale cutoff case, but not by much more than
the difference between the results of two different calculations in the
conventionl approach.

     The present derivation has the advantage that the zeroth order
approximation, the value of $\frac{M}{m}$ obtained for the vanishing of
the one-loop $\br$ at the weak scale, differs by less than ten percent
   from the final value for the large cutoff limit. If one makes the
corresponding zeroth order determination of the minimum M in the
   conventional approach by setting
$\beta_{\lambda}$ to zero at the weak scale, the result differs from
the final large cutoff result by more than thirty percent. The significant
difference is that $\br$ contains the large QCD correction to m, while
$\beta_{\lambda}$ does not. It is added in later as a correction when
one integrates the coupled RG equations.

It is clearly desirable to have a large scale lattice simulation study of the
combined Higgs-heavy quark-QCD sector. (Contributions from light quarks and
electroweak gauge bosons are small, particularly if one doesn't run up to some
very high scale).  We note that a quenched approximation
simulation is not adequate for this problem. The term in (\ref{deriv})
which triggers the possible instability is the $-2 N_c$, clearly a contribution
 from an internal fermion closed loop.


\begin{thebibliography}{10}

\bibitem{Sher} M.Sher,Phys.Rep. 179 (1989) 273; J.Gunion,H.Haber,
G.Kane,S.Dawson ''The Higgs Hunter's Guide'',Addison-Wesley (1990);
M.B.Einhorn,
''The Standard Model Higgs Boson'',North-Holland (1991).
\bibitem{AI} M.Sher,Phys.Lett.B 331(1994) 448; G.Altarelli,G.Isidori,
Phys.Lett. B 357(1994) 141.
\bibitem{Lin} L.Lin,I.Montvay,G.Munster,H.Wittig,Nucl.Phys.B 355(1991) 511.
\bibitem{BW1} A.I.Bochkarev,R.S.Willey,Phys.Rev.D 51(1995) R2049'
\bibitem{BW} A.I.Bochkarev,R.S.Willey, to be published

\end{thebibliography}
\end{document}